\documentclass[aps,showpacs,prb,reprint,superscriptaddress,longbibliography]{revtex4-1}

\pdfoutput=1

\usepackage{setspace} 
\usepackage{graphicx}
\usepackage{amsmath}
\usepackage{color}
\usepackage{amssymb}
\usepackage{verbatim}
\usepackage{latexsym}
\usepackage{enumerate} 
\usepackage{bm} 
\usepackage{subfigure}
\usepackage{mathtools}
\usepackage{braket}

\begin{document}

\title{First principles study of the dynamic Jahn-Teller distortion of
  the neutral vacancy in diamond}

\author{Joseph C.\ A.\ Prentice} 
\affiliation{TCM Group, Cavendish Laboratory, University of Cambridge,
  J.\ J.\ Thomson Avenue, Cambridge CB3 0HE, United Kingdom}
\author{Bartomeu Monserrat}
\affiliation{TCM Group, Cavendish Laboratory, University of Cambridge,
  J.\ J.\ Thomson Avenue, Cambridge CB3 0HE, United Kingdom}
\affiliation{Department of Physics and Astronomy, Rutgers University,
  Piscataway, New Jersey 08854-8019, USA}
\author{R.\ J.\ Needs} 
\affiliation{TCM Group, Cavendish Laboratory, University of Cambridge,
  J.\ J.\ Thomson Avenue, Cambridge CB3 0HE, United Kingdom}

\date{\today}

\begin{abstract}

  First-principles density functional theory methods are used to
  investigate the structure, energetics, and vibrational motions of
  the neutral vacancy defect in diamond.  The measured optical
  absorption spectrum demonstrates that the tetrahedral $T_d$ point
  group symmetry of pristine diamond is maintained when a vacancy
  defect is present. This is shown to arise from the presence of a
  dynamic Jahn-Teller distortion that is stabilised by large
  vibrational anharmonicity.  Our calculations further demonstrate
  that the dynamic Jahn-Teller-distorted structure of $T_d$ symmetry
  is lower in energy than the static Jahn-Teller distorted tetragonal
  $D_{2d}$ vacancy defect, in agreement with experimental
  observations.  The tetrahedral vacancy structure becomes more stable
  with respect to the tetragonal structure by increasing temperature.
  The large anharmonicity arises mainly from quartic vibrations, and
  is associated with a saddle point of the Born-Oppenheimer surface
  and a minimum in the free energy.  This study demonstrates that the
  behaviour of Jahn-Teller distortions of point defects can be
  calculated accurately using anharmonic vibrational methods.  Our
  work will open the way for first-principles treatments of dynamic
  Jahn-Teller systems in condensed matter.
\end{abstract}

\maketitle

\section{Introduction}

Point defects in crystals introduce electron energy levels within the
electronic band gap, trap charge carriers, emit and absorb light, and
phonon scattering from them limits thermal and electrical
conductivities.\cite{stoneham_theory_1975,schroder_quantum_2016} Such
effects influence many of the most desirable properties of diamond,
including its optical properties and high thermal conductivity, and
point defects in diamond have been the subject of much previous
theoretical
work.\cite{freysoldt_first-principles_2014,drabold_theory_2007} Some
defects such as the Si-V\cite{rogers_electronic_2014,gali_ab_2013} and
N-V$^-$ centres\cite{maurer_room-temperature_2012} have been
identified as potential ``qubits'' in quantum
computers.\cite{rogers_all-optical_2014,sipahigil_indistinguishable_2014,balasubramanian_ultralong_2009,bernien_heralded_2013,knowles_observing_2014,dolde_room-temperature_2013,schroder_quantum_2016}
Both these defects include a lattice vacancy, which is an important
defect in its own right. The neutral vacancy in diamond is
particularly significant because it is known to be stable over a wide
range of doping levels,\cite{corsetti_system-size_2011} and plays a
central role in defect diffusion.\cite{hood_quantum_2003} It is also
important for its optical absorption and luminescence properties -- it
is associated with a series of lines in the absorption spectrum of
diamond, including the strong and sharp GR1 line at
$1.673$~eV\cite{walker_optical_1979} -- and its applications in
quantum information\cite{wrachtrup_processing_2006} and precision
sensing.\cite{schirhagl_nitrogen-vacancy_2014}

\begin{figure*}
\begin{center}
\subfigure[][]{
\includegraphics[width=0.3\textwidth]{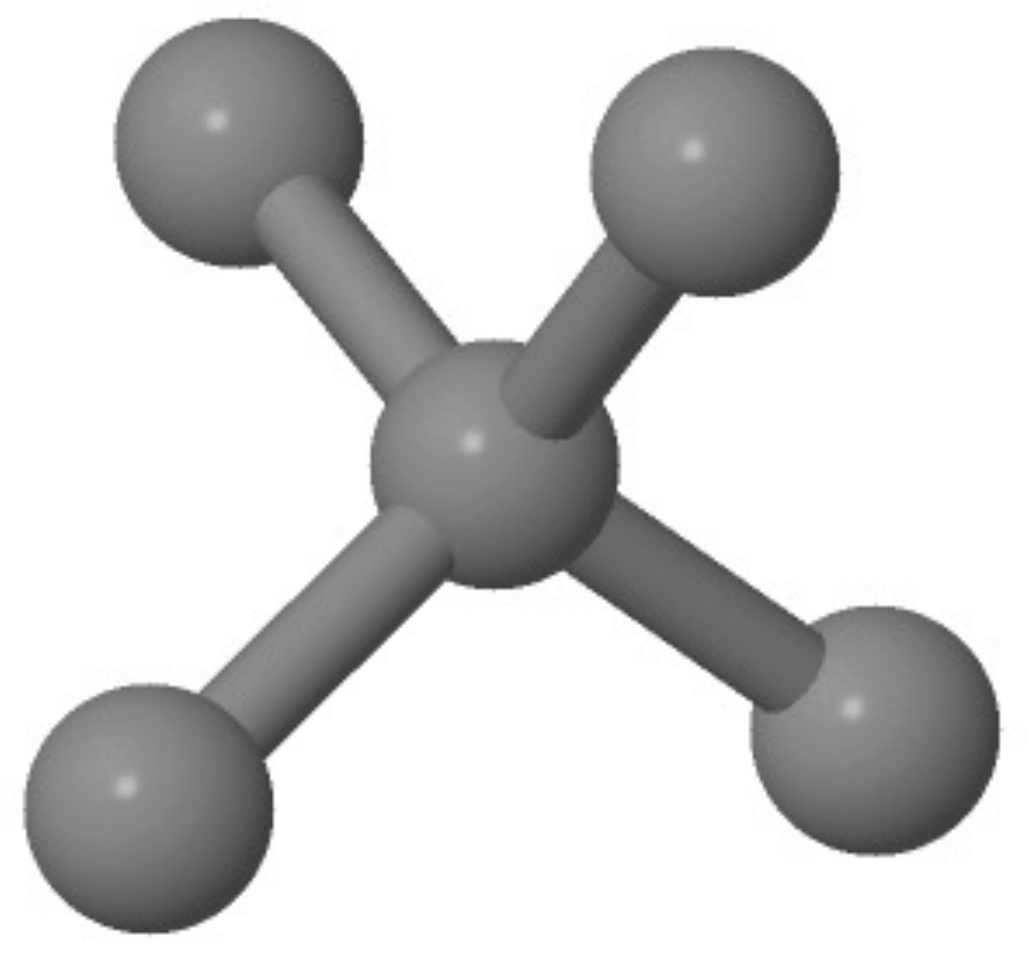}
\label{subfig:TetrahedralCoord}
}
~
\subfigure[][]{
\includegraphics[width=0.3\textwidth]{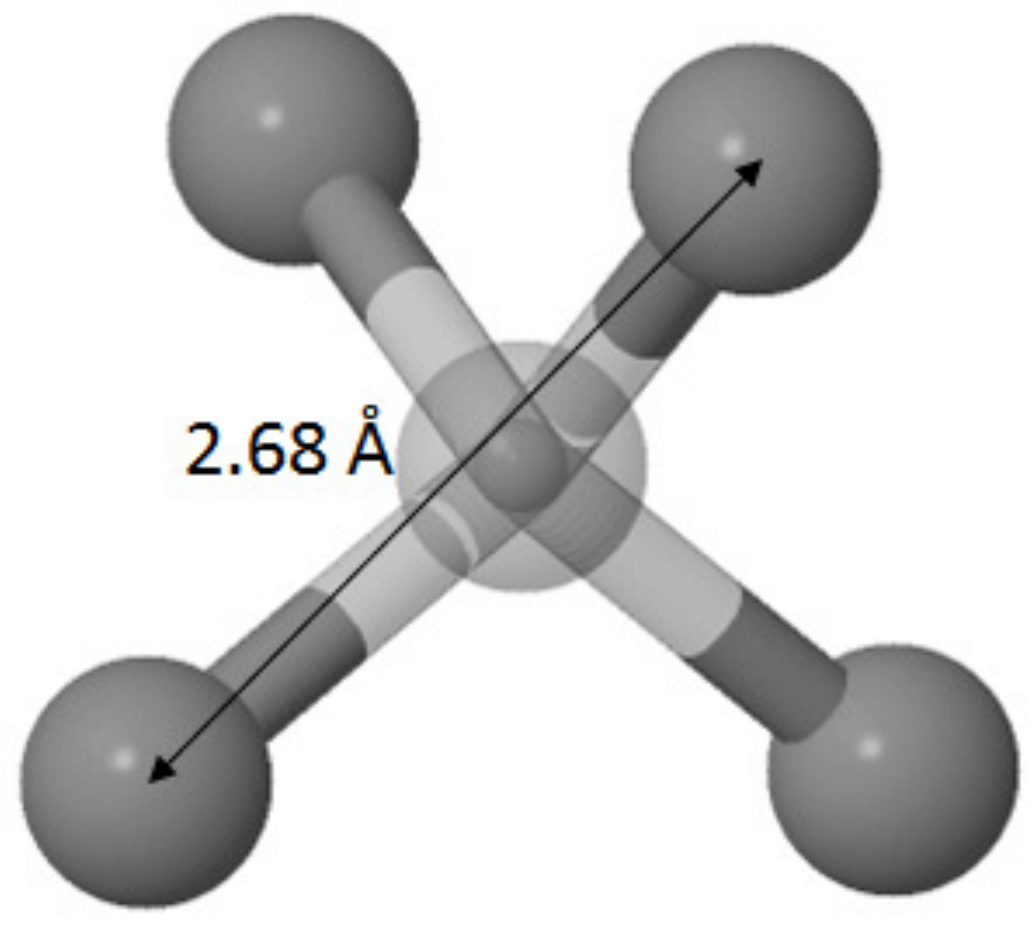}
\label{subfig:TdSymm}
}
~
\subfigure[][]{
\includegraphics[width=0.3\textwidth]{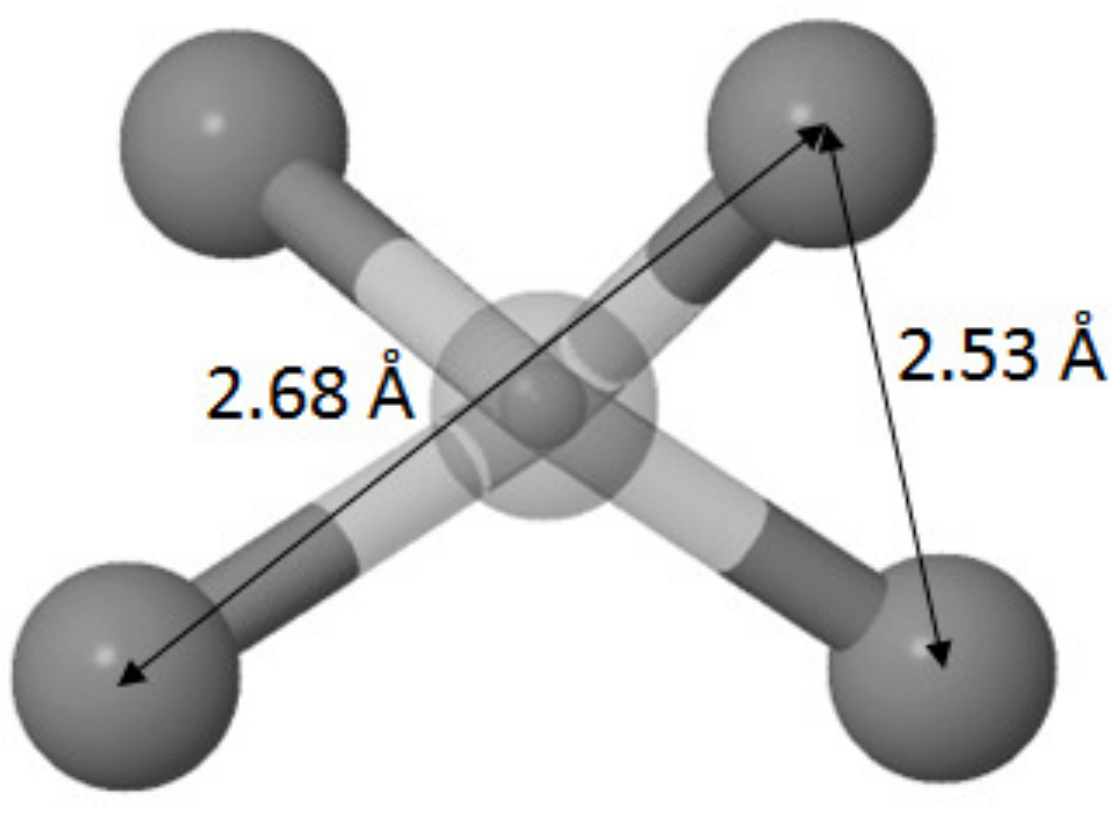}
\label{subfig:D2dSymm}
}
\end{center}
\caption{Structure of the pristine diamond lattice and possible
  distortions of the vacancy. (a) shows a site in the pristine
  lattice and its four nearest neighbours. (b) shows the nearest
  neighbours of the vacancy with a distortion of $T_d$ symmetry, and
  (c) shows the vacancy structure with $D_{2d}$ symmetry. The lengths 
  indicate the distances
  between atoms in the relaxed structures.  The four atoms in (c) form
  two pairs.} \label{fig:Symmetries}
\end{figure*}

Carbon and silicon are isoelectronic, and their pristine lattices have
tetrahedral $T_d$ point group symmetry, as depicted in Fig.\
\ref{subfig:TetrahedralCoord}. Watkins modelled the neutral vacancy in
silicon using a linear combination of atomic orbitals including only
the four atoms surrounding the vacancy,\cite{watkins_lattice_1986}
showing that two electrons should occupy a triply-degenerate
state. The energy of these electrons can be lowered by splitting this
degenerate energy state via a static Jahn-Teller
distortion\cite{jahn_stability_1937,watkins_lattice_1986} of
tetragonal $D_{2d}$
symmetry.\cite{corsetti_system-size_2011,probert_improving_2003,breuer_ab_1995}
Theoretically, the presence of a Jahn-Teller distortion is revealed by
the emergence in the undistorted structure of harmonic vibrational
soft modes, i.e., modes with imaginary frequencies; this approach
successfully predicts the $D_{2d}$ distortion experimentally observed
in the silicon vacancy. When applied to diamond, this approach also
predicts a $D_{2d}$ distortion, but in this case the experimental
observations show that the neutral diamond vacancy has $T_d$ symmetry
instead.\cite{davies_jahn-teller_1981,clark_neutral_1973} (Vacancy
structures with both symmetries are shown in Figs.\
\ref{subfig:TdSymm} and \ref{subfig:D2dSymm}.)  The experimental
observations can be rationalized by the appearance of a {\it dynamic}
Jahn-Teller
distortion,\cite{davies_jahn-teller_1981,lannoo_optical_1968,breuer_ab_1995,ham_jahn-teller_1971}
due to strong anharmonic vibrational motion. The dynamic Jahn-Teller
effect is observed in a variety of systems, including doped
manganites,\cite{millis_dynamic_1996,dediu_jahn-teller_2000}
fullerides,\cite{brouet_role_2001,canton_experimental_2002} octahedral
complexes of d$^9$ ions,\cite{o-brien_dynamic_1964} and the excited
states of the N-V$^-$ centre in diamond.\cite{fu_observation_2009}

In a dynamic Jahn-Teller system, there are two or more minima in the
Born-Oppenheimer (BO) energy surface, which are separated by energy
barriers. In the vacancy in diamond, there are three minima,
corresponding to tetragonal distortions in each Cartesian
direction. If the energy barriers between the minima are low enough,
the wavefunction of the system is shared between the minima instead of
localising in a single minimum, resulting in a dynamic Jahn-Teller
effect. This means that the vibrational wavefunction possesses the
symmetry of the undistorted structure.\cite{ham_jahn-teller_1971}
Structures at saddle points or maxima of the BO surface can be
stabilised by anharmonic vibrations, even at zero temperature, through
zero-point motion.  Previous estimates, from experimental data, of the
energy barriers between minima and an Einstein-like frequency for the
tetragonal defect modes of the vacancy in diamond show that the
vibrational energy quantum is larger than or comparable to the barrier
energy, implying a dynamic Jahn-Teller effect close to
$0$~K.\cite{stoneham_low-lying_1977}

The dynamic Jahn-Teller effect in the neutral vacancy in diamond is
well-established experimentally, but from a first-principles
standpoint the view is less clear. The commonly used harmonic
approximation for lattice dynamics cannot account for the presence of
a dynamic Jahn-Teller distortion, as this is an intrinsically
anharmonic effect. First-principles calculations of the anharmonic
vibrational wave function of the ground state of the neutral vacancy
in diamond have not been reported previously, meaning that a proper
description of the dynamic Jahn-Teller effect in this system is
lacking. In this study, we determine for the first time an anharmonic
wave function which accurately describes the dynamic Jahn-Teller
distortion, providing a theoretical description of the experimentally
observed $T_d$ symmetry of the neutral vacancy in diamond.

We have used first-principles DFT methods\cite{jones_density_2015} to
calculate the electronic and vibrational properties of the neutral
vacancy. The electronic ground state is expected to be well-described
within DFT, although some of its electronic excited states are not
because they have many-body (multideterminant) character. Our study is
confined to the electronic ground state.  The vibrational calculations
were carried out using a recently proposed vibrational self-consistent
field (VSCF) method, \cite{monserrat_anharmonic_2013} which neglects
non-adiabatic effects but includes anharmonicity.

We find that formation of the dynamic Jahn-Teller $T_d$ vacancy
structure leads to a small reduction in the vibrational energy
compared with the static Jahn-Teller $D_{2d}$ structure when
anharmonic effects are included. Our results show that anharmonic
nuclear motion leads to a dynamic Jahn-Teller distortion of the
vacancy in diamond very close to zero temperature, and we find that
this remains true at least up to $400$~K.

The rest of this work is organised as follows: in
Sec.~\ref{sec:methods} we outline the computational methods employed
and give technical details of the calculations. In
Sec.~\ref{sec:results} we present the main results of our study,
detailing the effects of including harmonic and anharmonic
contributions to the total energy, and to the vibrational density of
states in each symmetry state. In Sec.~\ref{sec:summary} we give a
brief summary of our results and some concluding remarks.

\section{Calculational Methods} \label{sec:methods}

\subsection{Electronic calculations}

DFT calculations were performed with version 7.0.3 of the {\sc castep}
code\cite{clark_first_2005} and the corresponding `on-the-fly'
ultrasoft carbon pseudopotential.\cite{vanderbilt_soft_1990} We have
used the local density approximation (LDA) as parametrised by Perdew
and Zunger,\cite{perdew_self-interaction_1981}, which has been widely
used in previous calculations involving diamond and similar
materials.\cite{al-mushadani_free-energy_2003,corsetti_system-size_2011,probert_improving_2003,kunc_equation_2003,maezono_equation_2007}
LDA-DFT calculations provide a lattice constant for
diamond\cite{jones_density_2015} of $3.529$~\r{A}, compared to the
experimental value of
$3.567$~\r{A}.\cite{madelung_semiconductors_1996} In our calculations,
the LDA-DFT lattice constant is generally used, although some
calculations were also carried out using the experimental lattice
constant to investigate the effect on the dynamic stability of the
tetrahedral symmetry state. Increasing the lattice constant was found
to increase the stability.

To ensure the existence of a locally stable tetragonal state (i.e.,
one that does not relax back to the tetrahedral symmetry when geometry
optimised), a 256-atom (255 when the vacancy is present) supercell is
used throughout this work.  In previous DFT work on the neutral
vacancy in silicon, it was observed that supercells of at least 256
atoms were required to obtain a stable tetragonal
distortion,\cite{corsetti_system-size_2011,probert_improving_2003}
constructed as a $2\times2\times2$ supercell of a 32-atom \textit{bcc}
unit cell.  Calculations with supercells with less than 256 atoms show
that the same is true for diamond.  The $256$-atom supercells also
allow us to effectively isolate the periodic defects from one another
so that accurate structures for the tetrahedral and tetragonal
vacancies can be obtained.

The geometry was optimised such that the root mean square of the
forces on all of the atoms was below $0.001$~eV/\r{A}. A simple 2-atom
unit cell was relaxed to obtain the relaxed LDA-DFT lattice constant,
and a 256-atom supercell was constructed from this. A vacancy was then
created and the internal co-ordinates relaxed.  The tetrahedral state
was found by imposing the pristine lattice symmetry before relaxing,
and the tetragonal state was found by imposing a tetragonal distortion
on the four nearest neighbours of the vacancy, before relaxing with no
symmetry constraints. In both vacancy structures, the nearest
neighbours relaxed away from the vacancy (by $0.11$~\r{A} in the
tetrahedral case) in order to increase their ${sp}^2$-like bonding ,
as observed in previous work.\cite{breuer_ab_1995}

A plane-wave cut-off energy of $650$~eV was used for relaxing the
structures and for the harmonic vibrational calculations, with a
$5\times5\times5$ Monkhorst-Pack $\mathbf{k}$-point
grid\cite{monkhorst_special_1976}, as the energy differences between
the structures were very well converged for these parameters. A larger
value of the cut-off energy was used for some calculations,
corresponding to even stricter convergence criteria. The SCF cycle
threshold for the energy was taken to be $10^{-10}$~eV per atom to
ensure a very accurate charge density, and thus accurate forces.

\subsection{Vibrational calculations}

A finite displacement method was used to obtain the matrix of force
constants, which was Fourier transformed to obtain the dynamical
matrix.\cite{kunc_ab_1982} This was diagonalised to obtain the
harmonic vibrational frequencies and eigenvectors.  Atomic
displacements of $0.00529$~\r{A} were used.  The anharmonic
vibrational calculations were conducted using the VSCF method
described in Ref.~\onlinecite{monserrat_anharmonic_2013} and used
successfully several times
since.\cite{azadi_dissociation_2014,monserrat_electron-phonon_2014,engel_anharmonic_2015}

In this method, the BO energy surface is described in a basis of
harmonic normal co-ordinates of amplitude $u$. In these coordinates,
the harmonic potential is separable, and each degree of freedom
contributes $\frac{1}{2}\omega^2u^2$, where $\omega$ is a harmonic
frequency. Starting from the harmonic approximation, we improve the
representation of the BO surface by using a principal axes
approximation,\cite{jung_vibrational_1996} which takes the form of a
sum over many-body terms, truncated at second order:
\begin{equation} 
  E(\mathbf{u})=E(\mathbf{0})+\sum_i V_1(u_i) + \frac{1}{2} \sum_{i'\neq i} V_2(u_i,u_{i'}).
\end{equation} 
In this expression, $\mathbf{u}$ is a vector containing the normal
mode amplitudes $u_i$, Cartesian indices are collective labels for the
quantum numbers $(\mathbf{q},\nu)$ where $\mathbf{q}$ is a phonon
wavevector and $\nu$ a phonon branch, and $E(\mathbf{u})$ is the value
of the BO energy surface when the atomic nuclei are in configuration
$\mathbf{u}$.  Anharmonicity is already included in the $V_1$ terms,
as they are not constrained to the harmonic form. The $V_2$ terms
provide additional anharmonic corrections arising from the
$2$-dimensional subspaces of the BO energy surface that they span. The
terms $V_1$ and $ V_2$ in this expansion are found by mapping the BO
energy surface as a function of the amplitude $u$ of each normal mode,
using a series of DFT calculations and then fitting cubic splines to
the results.  In this work, 19 different amplitudes per mode are used
in mapping the BO surface. Mapping the BO surface is by far the most
computationally expensive part of the entire calculation -- for
example, mapping the $V_1$ terms scales as the number of modes
multiplied by the number of mapping points $n$ multiplied by the cost
of a DFT calculation, which scales asymptotically as $nN^4$ for a
simulation cell with $N$ atoms.  The resulting anharmonic nuclear
Schr\"{o}dinger equation is solved using the VSCF method to give the
anharmonic vibrational energy and free energy.  The anharmonic
vibrational wavefunction $\ket{\Phi(\mathbf{u})}$ is written as a
Hartree product of the normal modes, $\prod_i \ket{\phi_i (u_i)}$. The
states $\ket{\phi_i (u_i)}$ are represented in terms of basis
functions which are the eigenstates of one-dimensional harmonic
oscillators. A total of $40$ basis functions were used for each degree
of freedom.

In most of this work only the $V_1$ terms are included in the
expansion of $E(\mathbf{u})$. The number of calculations required to
map the BO surface accurately as a function of two normal co-ordinates
is approximately the square of the number required for one
co-ordinate, making this computationally unfeasible if all pairs of
modes were to be included. Our tests show that the $V_2$ terms are
small for most modes, as the harmonic approximation works well for
vibrations in diamond.\cite{monserrat_anharmonic_2013} The only large
$V_2$ term corresponds to the $2$-dimensional BO subspace spanned by
two soft modes $(u_1,u_2)$ that are present in the tetrahedral
structure of the diamond vacancy. This $V_2$ term has three equivalent
minima, corresponding to a tetragonal distortion along each Cartesian
direction.  No matter how the orthogonal normal co-ordinates
$(u_1,u_2)$ are chosen, it is not possible for the axes to pass
precisely through more than one of these minima. This means that it is
impossible to capture the behaviour of the system with $3$ minima
using only the $V_1$ terms for these soft modes -- the associated
$V_2$ term must therefore be included.  Furthermore, the vibrational
wavefunction for this $V_2$ subspace of the BO surface cannot be
described correctly as a product of two states labelled by the two
corresponding normal modes $(u_1,u_2)$. Separating the wavefunction in
this way breaks the rotational symmetry of the problem.  Instead, for
this subspace, we separate the wavefunction using polar co-ordinates,
$r_u = \sqrt{u_1^2+u_2^2}$ and $\theta_u = \arctan(u_2/u_1)$, where
the $u_i$ are written in units of $1/\sqrt{2|\omega_{i,s}|}$, with
$\omega_{i,s}$ the imaginary harmonic soft phonon frequencies. This
preserves the correct symmetry of the problem and allows an accurate
wavefunction to be determined.  The usual wavefunction
$\ket{\phi_{1} (u_{1})}\ket{\phi_{2} (u_{2})}$ is re-expressed as
$\ket{R(r_u)}\ket{T(\theta_u)}$, where $\ket{R(r_u)}$ is written as a
sum of isotropic harmonic oscillator radial basis states, and
$\ket{T(\theta_u)}$ is a sum of sinusoids. A total of $80$ angular
basis functions and $20$ radial basis functions are used for these
calculations.  As all of the other modes are almost harmonic and
therefore well-described by the $V_1$ terms, the energy contribution
from this particular $V_2$ subspace can simply be added to the energy
of the rest of the $V_1$ terms of the tetrahedral defect.

It would be extremely computationally expensive to map all $762$
anharmonic modes for both the tetrahedral and the tetragonal symmetry
states of the vacancy, even if only the $V_1$ terms were
included. However, we can take advantage of the small anharmonicity in
pristine diamond.\cite{monserrat_anharmonic_2013} The fact that
pristine diamond is well-described by the harmonic approximation
implies that a necessary condition for modes to have significant
anharmonic character is that they are strongly affected by the
presence of the vacancy. We expect that these modes will generally
have short wavelengths and high energies, as over longer length scales
the effect of the vacancy will be averaged out. The effect of the
vacancy on each mode can be calculated as the difference between the
vacancy harmonic vibrational density of states (vDoS) and the pristine
vDoS at the frequency of each mode,
$\Delta g (\omega)=g_\text{vib}^\text{vac} (\omega) -
g_\text{vib}^\text{pris} (\omega)$.
We can then obtain an accurate value for the anharmonic correction to
the energy without having to map all 762 modes by simply mapping the
modes in descending order of $\Delta g (\omega)$, using the harmonic
approximation for all unmapped modes, and converging the anharmonic
correction with respect to the number of modes mapped. Converging the
correction to within $0.1$~meV requires 32 modes to be mapped for the
tetrahedral state, but 350 for the tetragonal state, implying that the
distortion away from tetrahedral symmetry leads to a large increase in
anharmonicity.

To further reduce the computational expense of the anharmonic
calculations, an energy cut-off of $350$~eV was used, with a
$5\times5\times5$ Monkhorst-Pack $\mathbf{k}$-point
grid\cite{monkhorst_special_1976}, as the shape of the BO surface is
well converged for these parameters. In some calculations larger
values of these parameters were used, corresponding to even stricter
convergence criteria. The energy differences between structures were
converged to within $0.5$~meV per atom, with the self-consistency
energy threshold for the DFT calculations set to $10^{-6}$~eV per
atom. The anharmonic correction to the energy in the pristine 256-atom
supercell was calculated from the anharmonic correction for a 16-atom
\textit{fcc} supercell.

\section{Results} \label{sec:results}

\subsection{Jahn-Teller effect and dynamical stability}

\begin{figure}
\begin{center}
\includegraphics[width=0.45\textwidth]{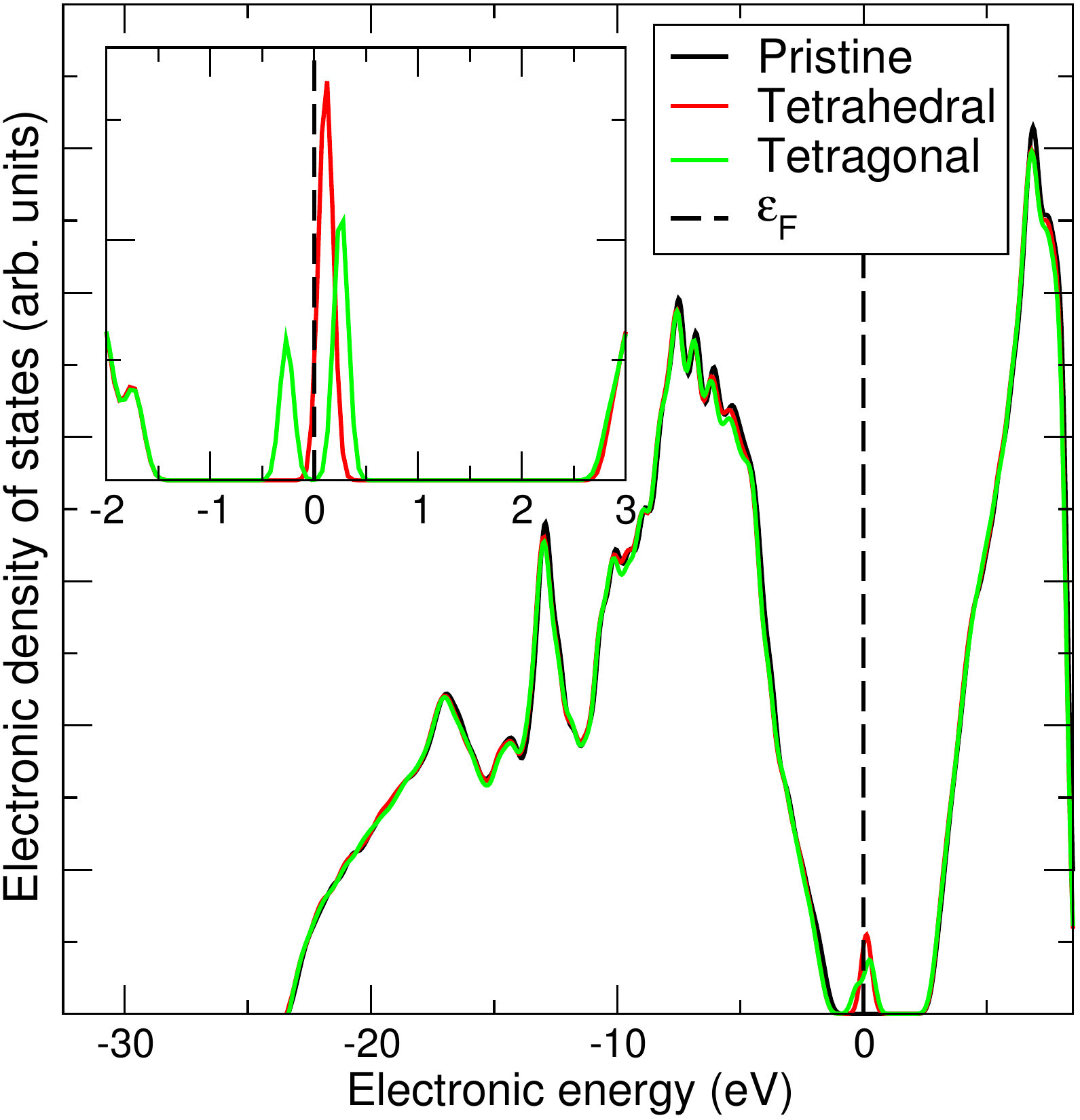}
\end{center}
\caption{Electronic density of states of diamond. The dotted line
  marks the Fermi energy, $\varepsilon_{\mathrm{F}}$. The inset contains the eDoS of the vacancy gap
  state, showing that the tetragonal distortion splits the state into a
  singlet at around $-0.3$~eV and a doublet at around $0.3$~eV.
} 
\label{fig:DiamondElecDoS}
\end{figure}

Figure \ref{fig:DiamondElecDoS} shows the calculated electronic
density of states (eDoS) for the pristine, tetrahedral and tetragonal
vacancy structures. There is little difference between the eDoS of the
pristine and vacancy states, apart from the appearance of a peak in
the band gap almost exactly at the Fermi level in the vacancy
structures. This state arises from the four `dangling bonds' around
the vacancy, as predicted by the Watkins model.  The inset to Fig.\
\ref{fig:DiamondElecDoS} shows that the peak splits into two upon
introduction of the tetragonal distortion, as would be expected for a
static Jahn-Teller distortion. These peaks correspond to the singlet
and doublet states predicted by the Watkins
model.\cite{watkins_lattice_1986}
Harmonic calculations show that the tetrahedral state is dynamically
unstable, that is, it exhibits two soft modes, with frequencies
$\omega_{1,s}$ and $\omega_{2,s}$, that correspond to tetragonal
distortions.  Unlike the tetrahedral symmetry state, the tetragonal
configuration is dynamically stable, leading to the conclusion that,
at this level of theory, a static Jahn-Teller distortion of tetragonal
symmetry is favoured.

\begin{figure}
\begin{center}
\includegraphics[width=0.47\textwidth]{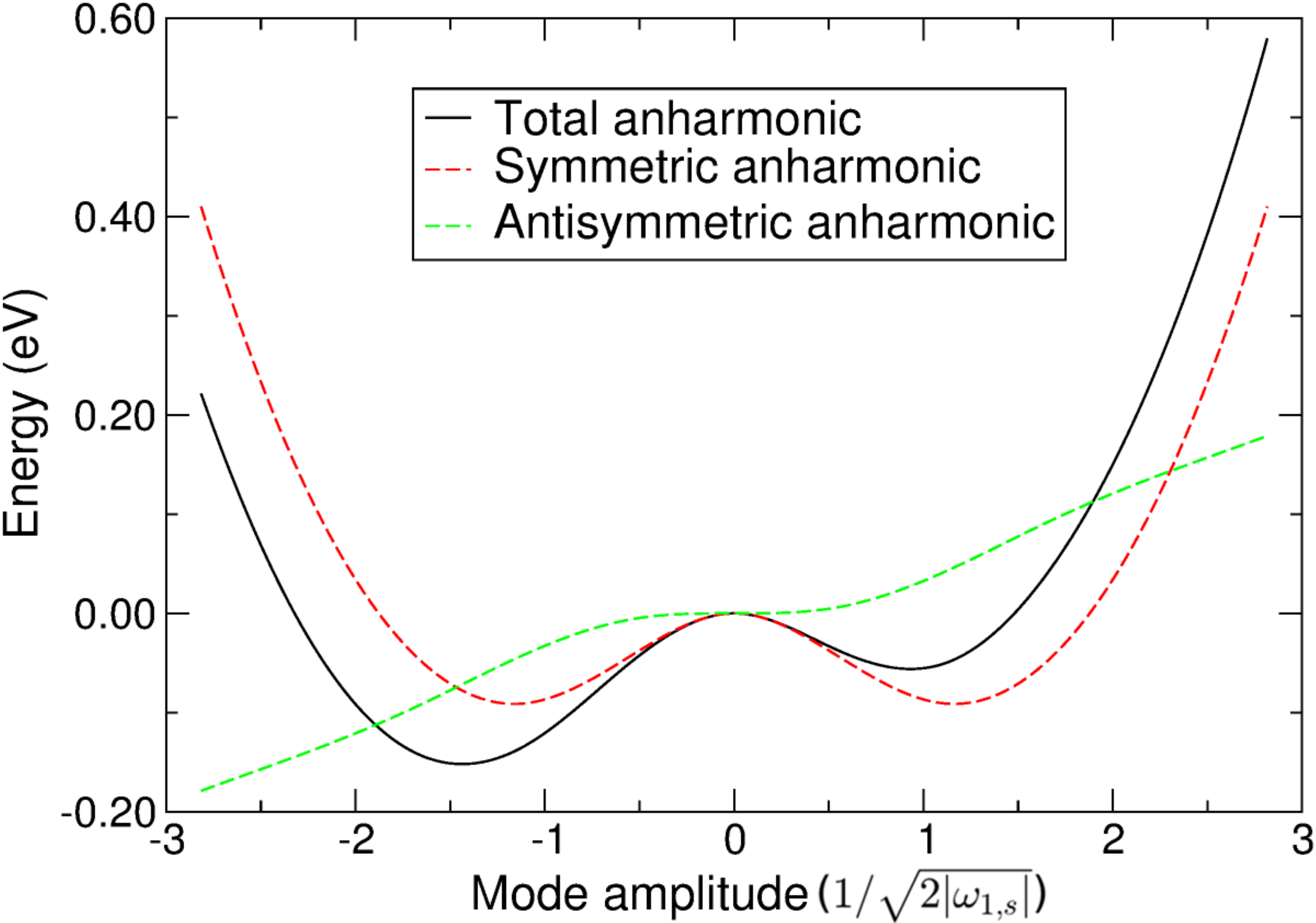}
\end{center}
\caption{Anharmonic Born-Oppenheimer energy surface mapped along the direction of one
  of the two soft modes of the tetrahedral structure, as well as its decomposition into
  symmetric and antisymmetric parts. The symmetric part of the anharmonic energy surface 
  is clearly quartic, giving two minima.}
\label{fig:QuarticAnh}
\end{figure}

\begin{figure}
\begin{center}
\subfigure{
\includegraphics[width=0.47\textwidth]{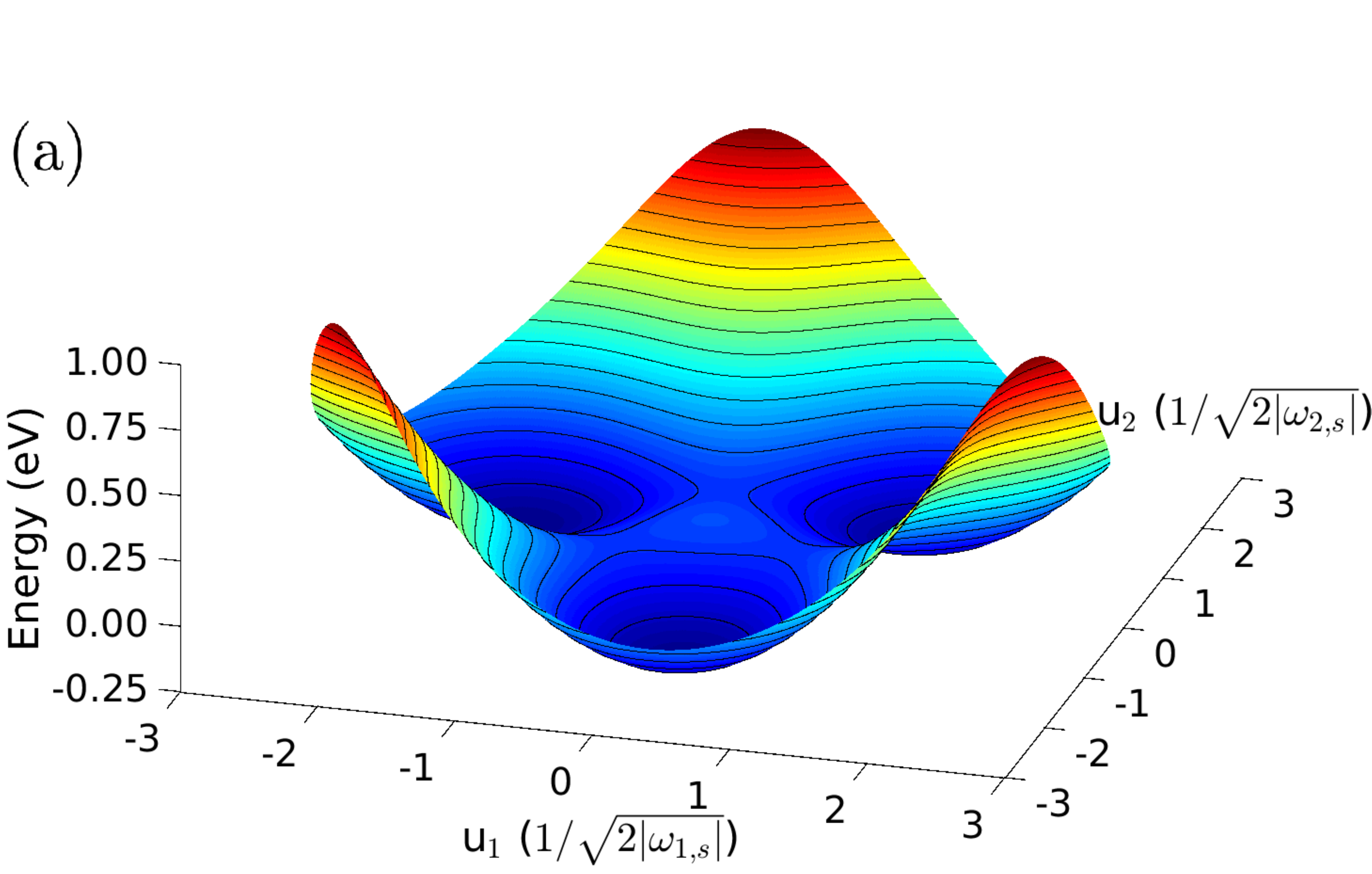}
\label{subfig:CoupledBOSurface}
}
\\
\subfigure{
\includegraphics[trim={4cm 2cm 0.75cm 4cm},clip,width=0.49\textwidth]{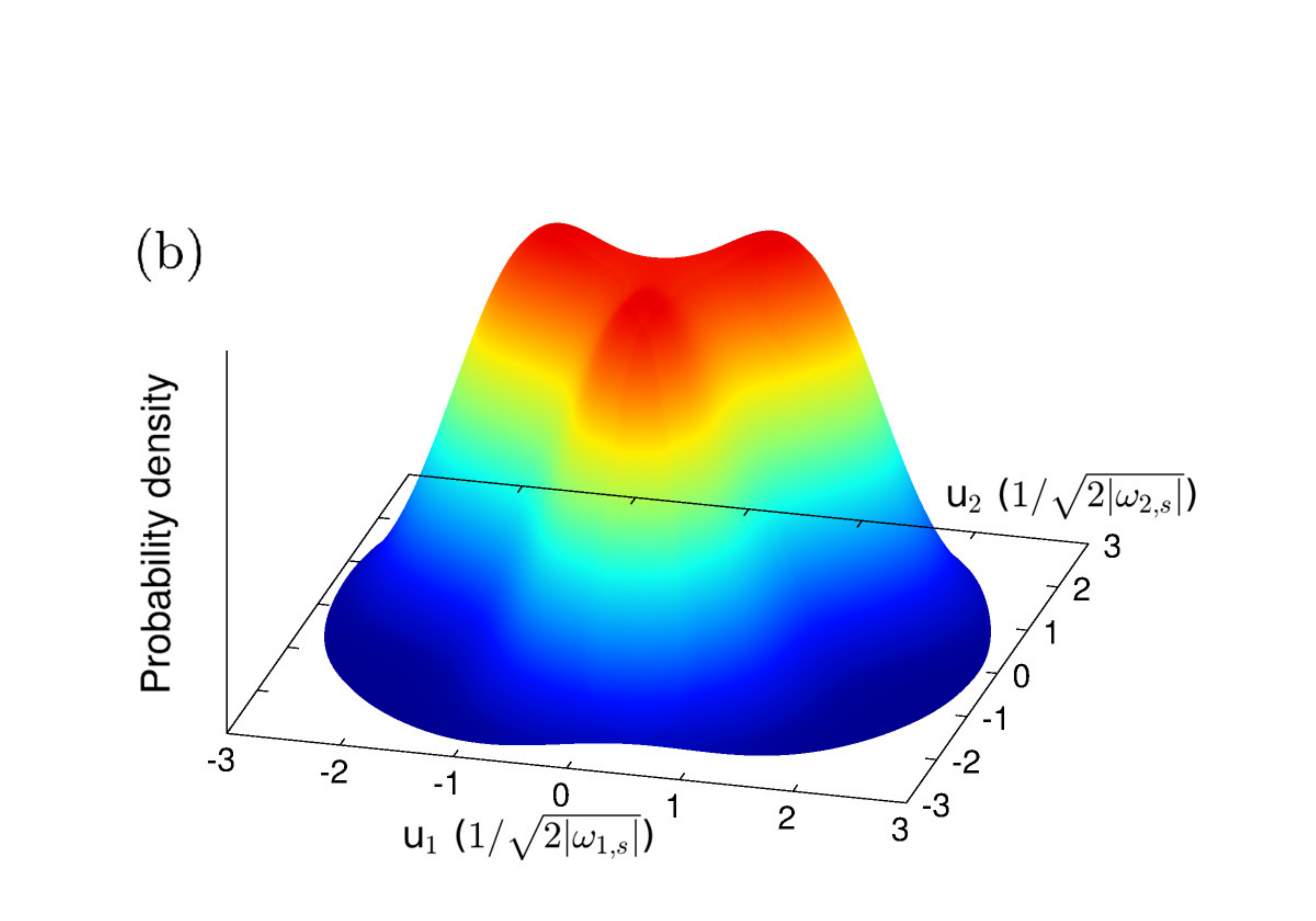}
\label{subfig:CoupledAnharmonicWavefn}
}
\end{center}
\caption{(a) Born-Oppenheimer energy surface in the plane spanned
  by the two soft modes of the tetrahedral structure. The
  static tetrahedral structure lies on the local maximum at the
  origin. Three equivalent minima corresponding to the three possible
  tetragonal distortions are arranged symmetrically around the
  tetrahedral structure. Each contour line represents an energy
  increase of $0.0615$~eV.\\
  (b) Anharmonic vibrational ground state probability density as a function
  of the amplitudes of the soft modes of the tetrahedral
  structure. The density has three peaks that correspond to the
  minima of the BO surface in (a).
 }
\label{fig:CoupledSurfWavefn}
\end{figure}

Including anharmonicity in the treatment of the tetrahedral
configuration provides a very different picture.  Figure
\ref{fig:QuarticAnh} shows the BO surface mapped along one of the soft
mode directions, split into symmetric and antisymmetric contributions
to the anharmonic energy surface. The antisymmetric part mostly arises
due to the fact that any given direction cannot pass through two
minima \textit{and} the origin, meaning that both minima cannot be
well mapped by a one-dimensional slice, as discussed above.  However,
the symmetric contribution clearly shows that quartic anharmonicity is
present.  Due to the impossibility of describing all three minima
correctly using one-dimensional subspaces spanned by either soft mode
individually, it is necessary to include the full $2$-dimensional
subspace spanned by both soft modes.  Figure
\ref{subfig:CoupledBOSurface} shows the $V_2$ term of the BO surface
mapped on the plane spanned by the two soft modes, and Fig.\
\ref{subfig:CoupledAnharmonicWavefn} shows the anharmonic ground state
nuclear density in this subspace. In addition to the slice shown in
Fig.\ \ref{fig:QuarticAnh}, further slices through the BO surface in
Fig.\ \ref{subfig:CoupledBOSurface} can be found in the Supplemental
Material.  The anharmonic ground state vibrational density of the
tetrahedral structure has peaks in each of the three minima of the BO
surface, which lowers the overall energy of the system. The fact that
the wavefunction is shared between the minima shows that, when
anharmonic vibrational effects are included, the Jahn-Teller effect in
this system becomes dynamic rather than static, with the system
maintaining the full $T_d$ point symmetry of the pristine lattice.

The dynamical stability of the tetrahedral state is somewhat sensitive
to the exact form of the BO surface found in the DFT calculations. If
the relaxed LDA lattice constant ($3.529$~\r{A}) is used when mapping
the $2$-dimensional BO surface of the pair of soft modes, the
tetrahedral state is still dynamically unstable at $0$~K, becoming
stable at $16.9$~K.  However, using the experimental lattice constant
of $3.567$~\r{A}\cite{madelung_semiconductors_1996} reduces the size
of the dynamical instability significantly, decreasing the absolute
value of the already small ground state energy associated with the
$V_2$ subspace by an order of magnitude.  Using the experimental
lattice constant the tetrahedral state is calculated to become
dynamically stable at $8.6$~K. Given the errors inherent in DFT
calculations, our results are consistent with the tetrahedral state
being dynamically stable down to liquid helium temperatures, as
implied by experiment,\cite{lannoo_optical_1968} and even to absolute
zero.

The minima in the BO surface are, in the polar co-ordinates defined
above, at $r_u=1.64/\sqrt{2|\omega_s|}$,
$\theta_u = 39^\circ,159^\circ,279^\circ$, which correspond to
tetragonal distortions along the $x$, $z$ and $y$ directions,
respectively.
$\omega_s=(\omega_{1,s}+\omega_{2,s})/2\simeq\omega_{1,s}\simeq\omega_{2,s}$.
The values of $\theta_u$ depend on the precise choice of the axes
defined by the two soft modes.  The displacement patterns
corresponding to the modes $u_1$ and $u_2$ in this work are presented
in the Supplemental Material, allowing these minima to be
unambiguously identified.  At these minima, the four nearest
neighbours of the vacancy are displaced from their tetrahedral
equilibrium positions; they are displaced by $0.074$~\r{A}
\textit{away} from the vacancy along the distortion direction, but by
half this distance \textit{towards} the vacancy in the other two
directions. The tetrahedral structure is at a maximum of the BO
surface along the direction of the soft modes, but a minimum along all
of the other modes, placing the tetrahedral structure at a saddle
point of the BO surface.

\subsection{Thermodynamics}

\begin{figure}[t!]
\begin{center}
\subfigure[][]{
\includegraphics[width=0.47\textwidth]{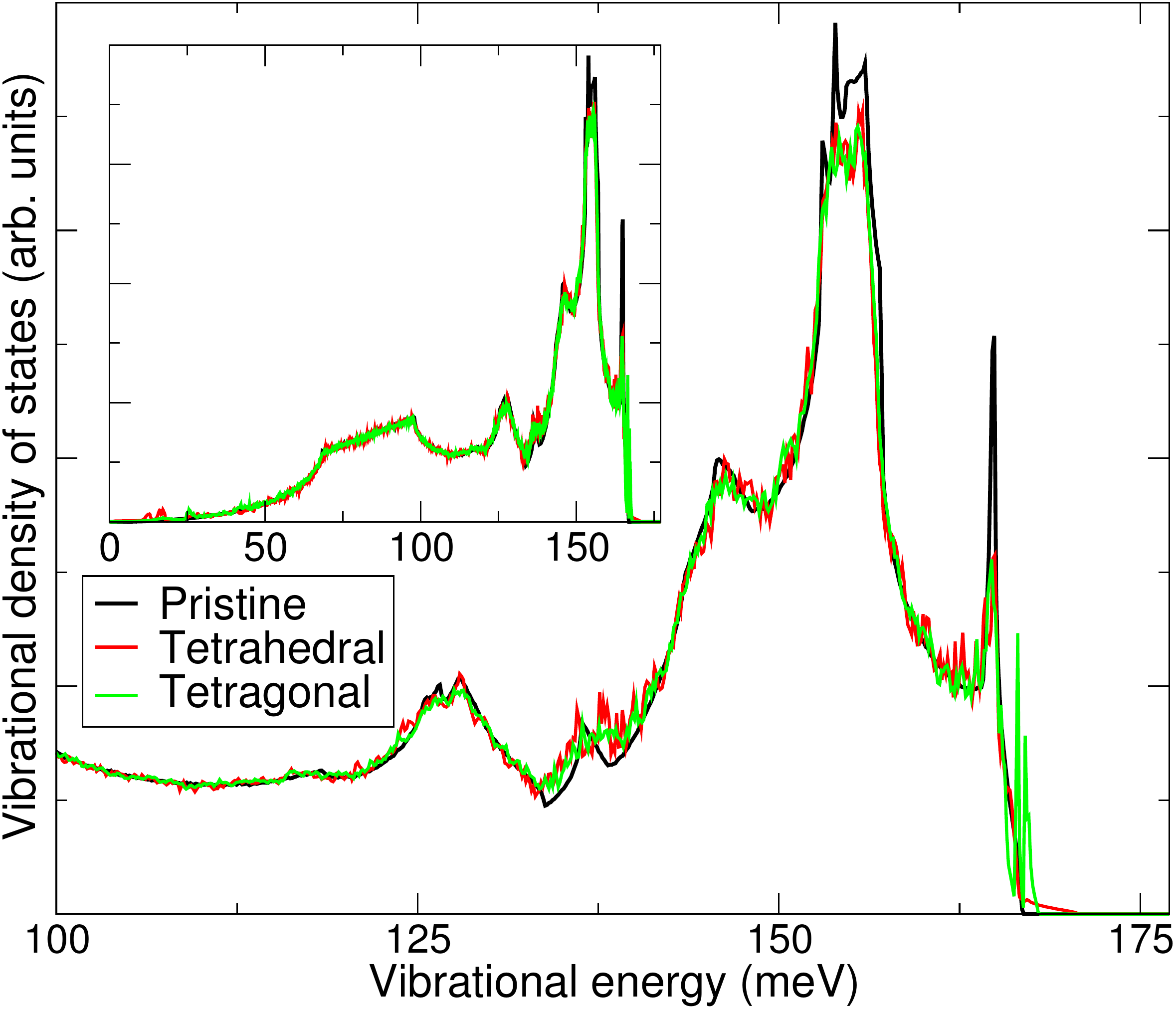}
\label{subfig:HarvDoS}
}
\\
\subfigure[][]{
\includegraphics[width=0.47\textwidth]{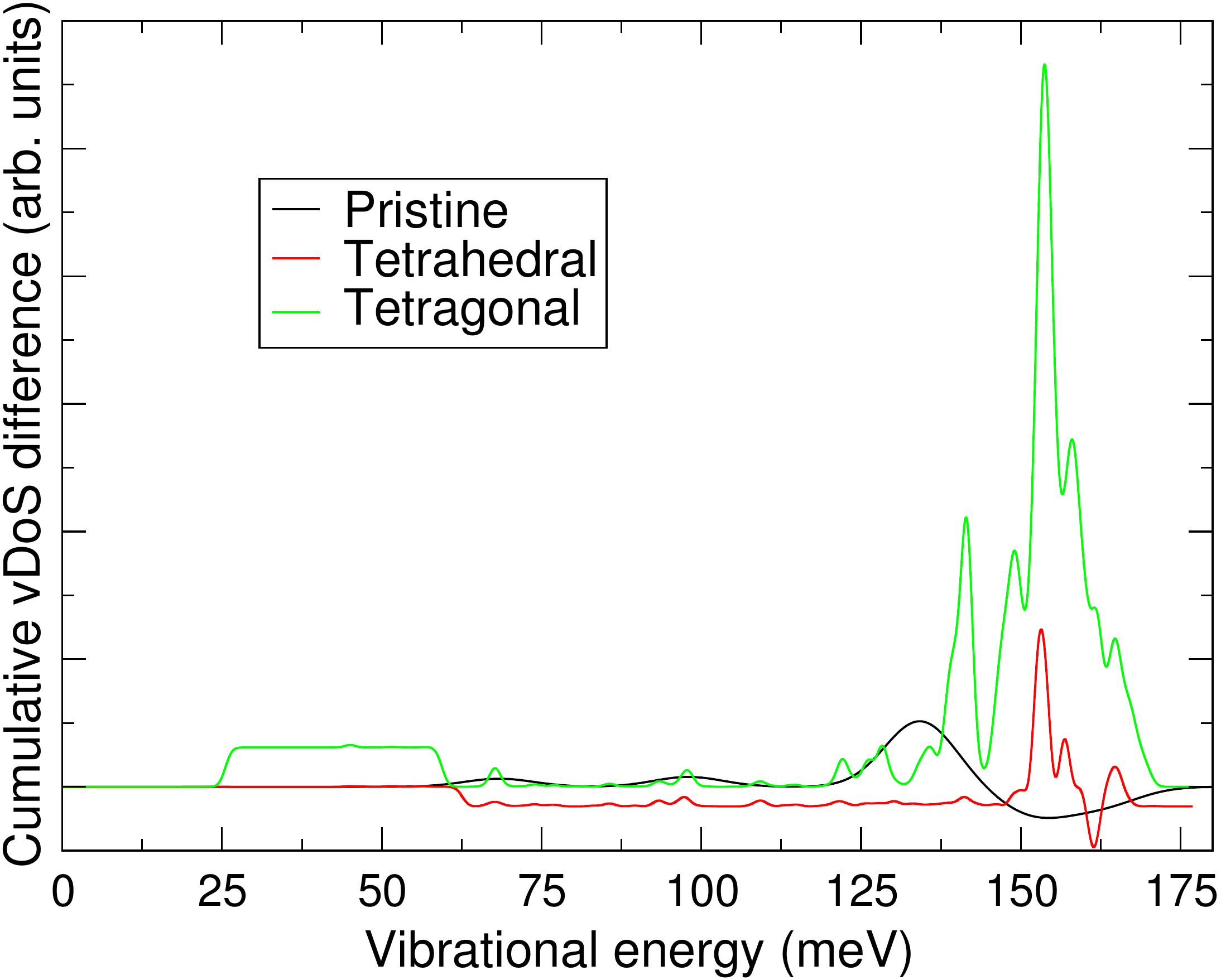}
\label{subfig:AnhvDoS}
}
\end{center}
\caption{(a) shows the harmonic vibrational density of states for the pristine,
  tetrahedral vacancy and tetragonal vacancy structures in diamond,
  shown at high energies above $0.1$~eV in the main plot, and in full in the
  inset. Note the main differences between the vacancy states and the
  pristine structure occur at high energies.\\
  (b) shows the difference between the harmonic and anharmonic cumulative vDoS, $\Delta G(\varepsilon)=\int^\varepsilon_0 d\varepsilon' g_{\text{har}}(\varepsilon')-g_{\text{anh}}(\varepsilon')$,
  of the vacancy and pristine structures, constructed
  using Gaussian smearing of the frequencies.
}
\end{figure}

The tetragonal and pristine structures are dynamically stable at the
harmonic level, and the previous section shows that the tetrahedral
structure is also dynamically stable at low temperatures when
anharmonicity is accounted for. Having therefore established that all
three structures -- tetrahedral, tetragonal and pristine -- are
dynamically stable at low temperature at the anharmonic level, we turn
to their thermodynamics.  The static lattice, vibrational, and
formation energies at $20$~K are reported in Table
\ref{tab:EfEharEanh} for all three structures, as at this temperature
all three are dynamically stable.  The harmonic energy $E_\text{har}$
and the anharmonic correction
$\Delta E_\text{anh}=E_\text{anh}-E_\text{har}$, per atom, are given,
as well as the vacancy formation energy, which is calculated
as:\cite{corsetti_system-size_2011}
\[ E_f = E_\text{vac} - \frac{N-1}{N} E_\text{pris}, \] where
$E_\text{vac}$ and $E_\text{pris}$ are the total energies of the
system with and without the vacancy, respectively, and $N$ is the
number of atoms in the pristine supercell. The values of $E_f$ for the
two different symmetry states of the vacancy are presented in the
third part of Table \ref{tab:EfEharEanh}, at three levels of theory --
static (electronic), harmonic vibrational, and anharmonic
vibrational.

Because the tetrahedral state is dynamically unstable at the harmonic
level (although not at the anharmonic level), due to the presence of
the two soft modes, a harmonic vibrational energy cannot strictly be
defined for this structure. Despite this, we have included an
estimated value for the tetrahedral harmonic energy, calculated by
simply cutting out the contribution of the two soft modes, to enable
comparisons between the two symmetry states. The fact that this value
involves the unphysical removal of two modes is noted in Table
\ref{tab:EfEharEanh}.

\begin{table*}[t]
\centering
\begin{tabular}{ c | c | c c | c c c | }
\cline{2-7}
 & \multicolumn{1}{ | c | }{Static energies} & \multicolumn{2}{ | c | }{Vibrational energies} & \multicolumn{3}{ | c | }{Vacancy formation energies} \\
\hline
\multicolumn{1}{| c |}{Structure} & \multicolumn{1}{| c |}{$E_\text{static}$ (eV/atom)} & $E_\text{har}$ (eV/atom) & $\Delta E_\text{anh}$ (meV/atom) & $E_f^\text{static}$ (eV) & $E_f^\text{har}$ (eV) & $E_f^\text{anh}$ (eV) \\
\hline
\multicolumn{1}{| c |}{Tetrahedral} & \multicolumn{1}{| c |}{0.0292} & (\emph{0.1826}) & 0.190 & 7.451 & (\emph{8.343}) & 8.373 \\
\multicolumn{1}{| c |}{Tetragonal} & \multicolumn{1}{| c |}{0.0281} & 0.1831 & 0.821 & 7.176 & 8.185 & 8.376 \\
\multicolumn{1}{| c |}{Pristine} & \multicolumn{1}{| c |}{0.0000} & 0.1791 & 0.071 & 0.000 & 0.000 & 0.000 \\
\hline
\end{tabular}
\caption{DFT static lattice energies, vibrational energies and formation energies for each structure at $20$~K. 
  The second column shows the electronic static lattice energy $E_\text{static}$ per atom for the 
  pristine, tetrahedral and tetragonal structures relative to the pristine structure. The next two columns 
  show the harmonic vibrational energy $E_\text{har}$ and anharmonic energy correction 
  $\Delta E_\text{anh}=E_\text{anh}-E_\text{har}$ per atom for the three structures. The last three columns 
  show the formation energy at the static, harmonic, and anharmonic levels of theory, $E_f^\text{static}$, 
  $E_f^\text{har}$ and $E_f^\text{anh}$, respectively. The tetrahedral structure is dynamically unstable 
  at the harmonic level, as marked by \emph{(italicised brackets)}.
}
\label{tab:EfEharEanh}
\end{table*}

With this caveat in mind, we can look at the thermodynamic stability
of the two symmetry states at $20$~K. When only electronic and
harmonic effects are included in the formation energy, the state with
tetragonal symmetry is the most stable, although the inclusion of
harmonic vibrational effects reduces the Jahn-Teller relaxation energy
-- the energy difference between the tetrahedral and tetragonal
structures -- from $0.275$~eV to $0.158$~eV.  Upon inclusion of
anharmonic effects, the tetrahedral state becomes the most stable, as
observed experimentally, by $3$~meV.  The predicted final formation
energy for the neutral vacancy, including anharmonic effects, is
therefore $8.373$~eV, which is close to the estimates from experiments
of $9$--$15$~eV.\cite{bourgoin_experimental_1983} The formation energy
of the unrelaxed vacancy at the static level is calculated to be
$8.166$~eV, implying a total relaxation energy of $0.989$~eV at this
level of theory. (This result is not included in Table
\ref{tab:EfEharEanh}).

Comparing the vDoS of the vacancy structures to that of pristine
diamond gives further insight into the effect of the vacancy on the
vibrational properties. Figure \ref{subfig:HarvDoS} shows the harmonic
vDoS for all three structures at high energies, with the full vDoS as
an inset, and Fig.\ \ref{subfig:AnhvDoS} shows the difference between
the harmonic and anharmonic cumulative vDoS,
$\Delta G(\omega)=\int^\omega_0 d\omega'
g_{\text{har}}(\omega')-g_{\text{anh}}(\omega')$,
for both symmetry states of the vacancy as well as the pristine
lattice.  The cumulative densities of states were formed by broadening
the mode frequencies with Gaussians (of width $8.163\times 10^{-4}$~eV
for the vacancy states and $5.442\times 10^{-3}$~eV for the pristine
lattice) and cumulatively summing them.  This allows us to see how the
presence of anharmonicity changes the frequencies themselves. Figure
\ref{subfig:HarvDoS} confirms that the presence of the vacancy only
has a significant effect on high energy vibrations. This justifies our
approach of including only the highest energy vibrational modes in the
anharmonic calculations. The atoms neighbouring the vacancy tend to
have larger vibrational amplitudes than the other atoms in the very
highest energy modes for both symmetry configurations, while for lower
energy modes the amplitudes are comparable.

Figure \ref{subfig:AnhvDoS} shows that the effect of anharmonicity is 
much more pronounced in the tetragonal configuration than in the 
tetrahedral or pristine structures. In the tetrahedral and pristine 
structures, the changes in the vDoS are of a similar size, and are much 
smaller than the changes seen in the tetragonal case. This demonstrates 
that distortions away from the tetrahedral symmetry of the 
pristine lattice strongly increase the anharmonicity of the phonon modes, 
with the tetrahedral vacancy retaining the weak anharmonic character 
of pristine diamond. The optical modes at high energies are clearly 
more affected by the inclusion of anharmonicity than the low energy 
acoustic modes. In the pristine and tetrahedral structures, 
anharmonicity raises the frequency of some modes whilst lowering those 
of others, leading to both positive and negative values of $\Delta G$. 
In the tetragonal configuration, however, it generally raises the 
frequency of the modes by a small amount, showing that the leading 
anharmonic term is quartic in character, as cubic anharmonicity always 
acts to lower the energy in one dimension.\cite{monserrat_anharmonic_2013}

\begin{figure}[th]
\centering
\includegraphics[width=0.47\textwidth]{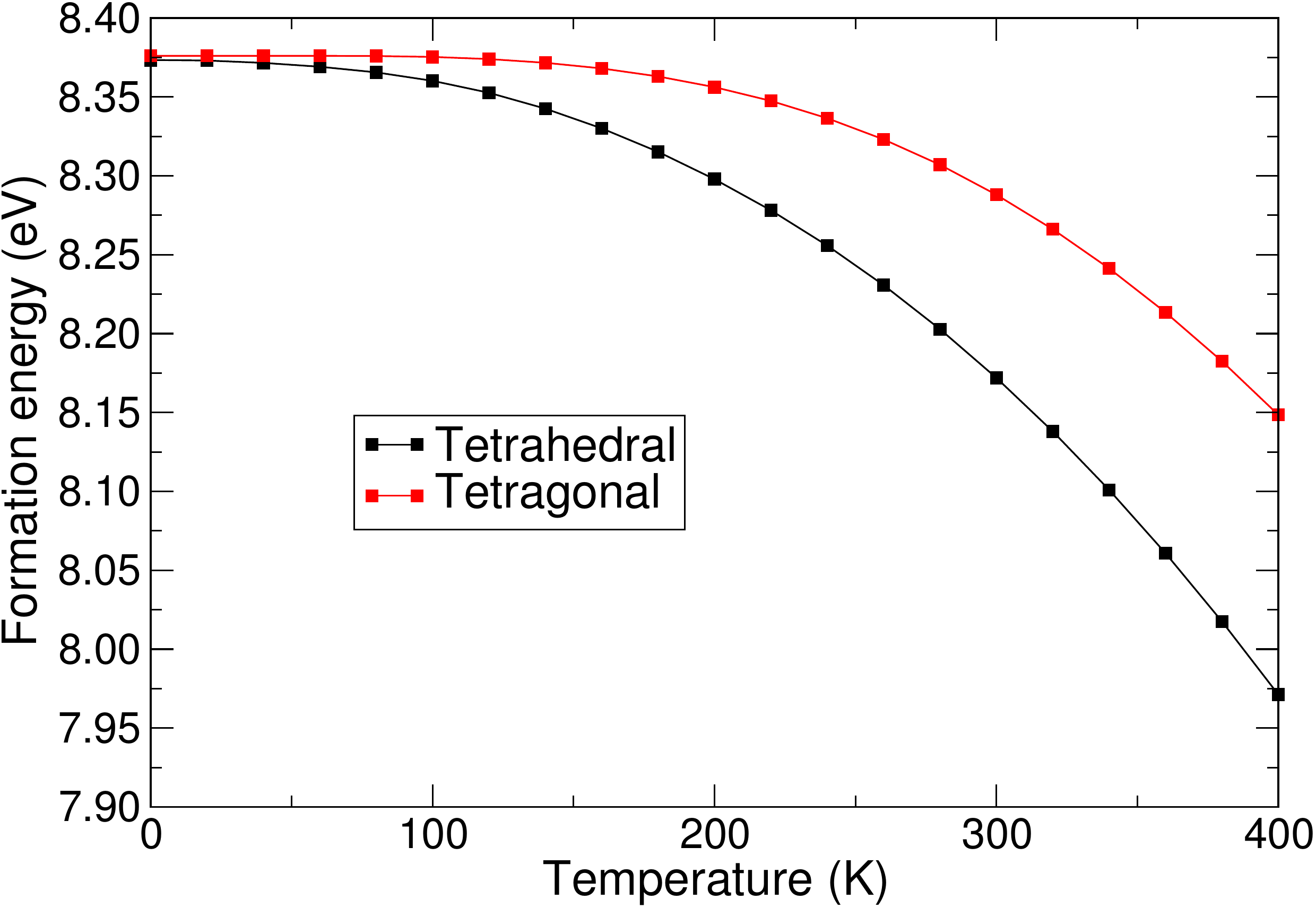}
\caption{Temperature dependence of the formation energies of the
  tetrahedral and tetragonal symmetry configurations of the neutral vacancy,
  including anharmonic effects but neglecting the small thermal expansion.
}
\label{fig:TDependence}
\end{figure}

Given the above results at $20$~K, we briefly examine the temperature
dependence of the formation energies of the tetrahedral and tetragonal
structures. Neglecting the effect of thermal expansion, which is very
small for diamond over the range of temperatures
considered,\cite{stoupin_thermal_2010} we calculate the anharmonic
vibrational contribution to the free energy at a set of finite
temperatures, using the excited states of the VSCF Hamiltonian to
construct a partition function $\mathcal{Z}$, from which we can
calculate a free energy
$F=-k_BT \ln \mathcal{Z}$.\cite{monserrat_anharmonic_2013} For these
calculations, 80 basis functions were used to obtain accurate excited
states. Figure \ref{fig:TDependence} shows the anharmonic formation
energy of each symmetry state of the vacancy for a range of
temperatures up to $400$~K. It is clear that the tetrahedral structure
remains the most stable over this temperature range -- indeed, the
difference in the formation energies of the two symmetry states
increases from $0.003$ to $0.177$~eV at $400$~K. The calculated value
of the vacancy formation energy at room temperature ($300$~K) is
$8.172$~eV, again in reasonable agreement with experimental estimates
of $9$--$15$~eV.\cite{bourgoin_experimental_1983}

\section{Conclusions} \label{sec:summary}

Our results show that the tetrahedral symmetry structure of the
neutral vacancy in diamond is stabilised down to almost zero
temperature by anharmonic vibrations.  The vacancy undergoes a dynamic
Jahn-Teller distortion which has the full $T_d$ point group symmetry
of the pristine system, as observed experimentally. The anharmonic
vibrational wavefunction of the tetrahedral defect has been
calculated, and shown to be shared evenly among the three minima in
the Born-Oppenheimer surface, which correspond to the three tetragonal
distortions. We have also calculated the temperature dependence of the
vacancy formation energy up to $400$~K. Our value for the formation
energy of the neutral vacancy agrees well with experimental estimates
of $9$--$15$~eV.\cite{bourgoin_experimental_1983}

Our results for the isolated neutral vacancy also imply that our
method could be used to calculate anharmonic properties of other point
defects, including those in diamond. Two examples of such defects are
the Si-V or N-V centres, which are especially of interest due to their
possible use as qubits in quantum
computing.\cite{rogers_all-optical_2014,sipahigil_indistinguishable_2014,balasubramanian_ultralong_2009,bernien_heralded_2013,knowles_observing_2014,dolde_room-temperature_2013,maurer_room-temperature_2012}
Studies of their vibrational properties, including anharmonicity,
would lead to a fuller understanding of these important defects.

\section{Acknowledgements}

J.C.A.P. and R.J.N. thank the Engineering and Physical Sciences
Research Council (EPSRC) of the UK for financial support
[EP/J017639/1]. B.M.  acknowledges Robinson College, Cambridge, and
the Cambridge Philosophical Society for a Henslow Research Fellowship.
Computational resources were provided by the High Performance
Computing Service at the University of Cambridge and the Archer
facility of the UK's national high-performance computing service, for
which access was obtained via the UKCP consortium [EP/K014560/1].

\bibliography{VacancyBib}

\end{document}


\title{Supplemental Material for ``First principles study of the dynamic Jahn-Teller distortion of the neutral vacancy in diamond''}

\author{Joseph C.\ A.\ Prentice}
\affiliation{TCM Group, Cavendish Laboratory, University of Cambridge,
  J.\ J.\ Thomson Avenue, Cambridge CB3 0HE, United Kingdom}
\author{Bartomeu Monserrat}
\affiliation{TCM Group, Cavendish Laboratory, University of Cambridge,
  J.\ J.\ Thomson Avenue, Cambridge CB3 0HE, United Kingdom}
\affiliation{Department of Physics and Astronomy, Rutgers University,
  Piscataway, New Jersey 08854-8019, USA}
\author{R.\ J.\ Needs}
\affiliation{TCM Group, Cavendish Laboratory, University of Cambridge,
  J.\ J.\ Thomson Avenue, Cambridge CB3 0HE, United Kingdom}

\date{\today}

\maketitle

\section{Pseudopotentials}

All density functional theory calculations were performed using {\sc CASTEP} version $7.0.3$, and its own ``on-the-fly'' ultrasoft pseudopotential for carbon. The definition string for the pseudopotential was:

\begin{verbatim}
2|1.4|1.4|1.3|6|10|12|20:21(qc=6)
\end{verbatim}

\section{Equilibrium positions}

The equilibrium positions of the 255 atoms in the tetrahedral symmetry structure with the experimental lattice constant are given in the file \texttt{TdSymmetry.cif}. The 4 nearest neighbours of the vacancy are the 1st, 37th, 41st and 69th atoms listed.

\section{Soft mode displacement patterns}

The displacement patterns of the two soft modes, labelled by $u_1$ and $u_2$ in the main text, are given in the files \texttt{u1Displacements.cif} and \texttt{u2Displacements.cif}. The 4 nearest neighbours of the vacancy are the 1st, 37th, 41st and 69th atoms listed. 

\section{Born-Oppenheimer energy surface slices}

\begin{figure}
\begin{center}
\subfigure[][]{
\includegraphics[width=0.75\textwidth]{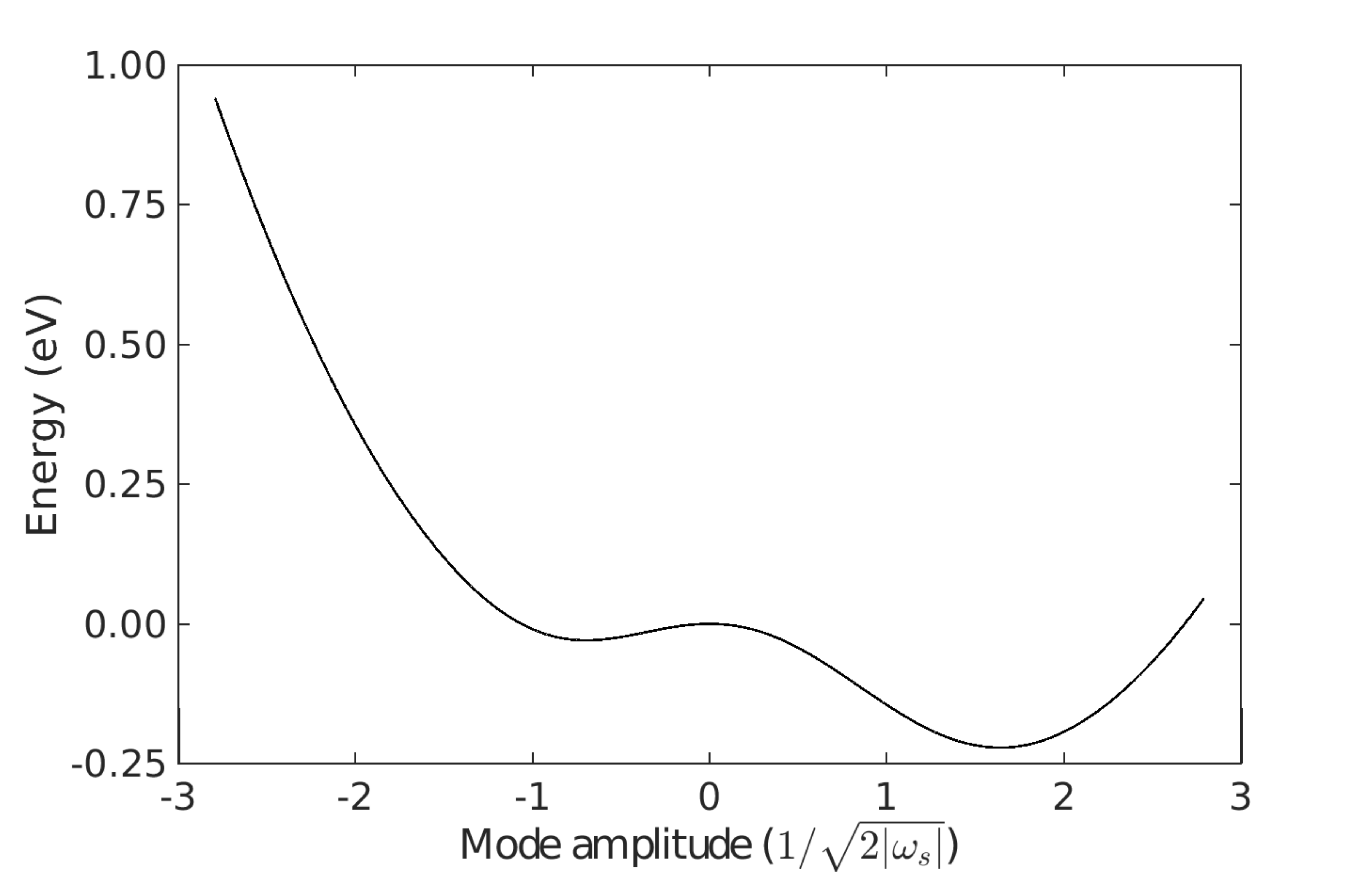}
\label{subfig:mintomax}
}
~
\subfigure[][]{
\includegraphics[width=0.75\textwidth]{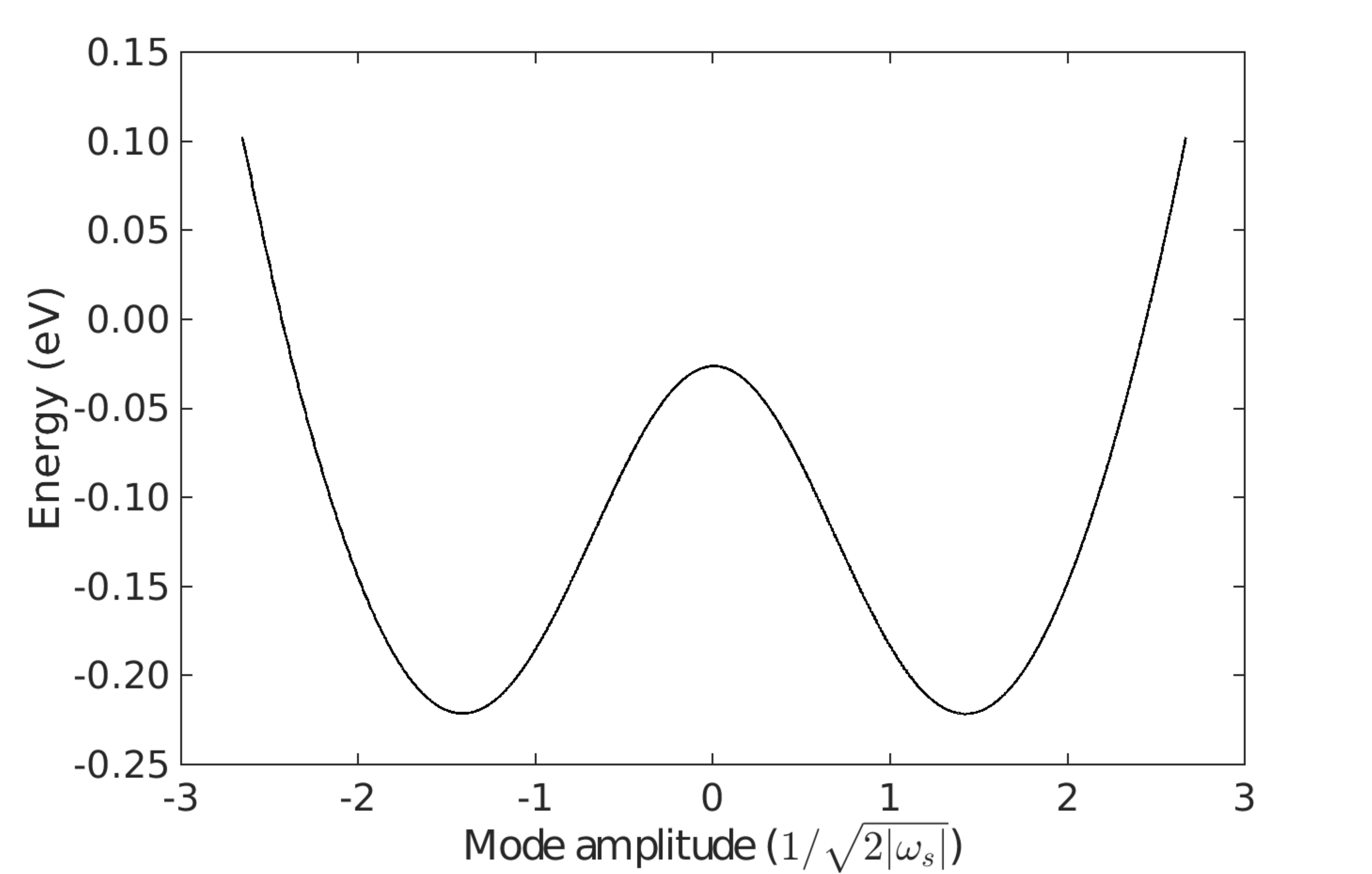}
\label{subfig:mintomin}
}
\end{center}
\caption{Slices through the BO surface in the plane spanned by the two soft modes of the tetrahedral structure. The slice in (a) runs through one minimum and the origin, also passing through the high point of the graph opposite. The slice in (b) runs through two adjacent minima. In (b), the point midway between the two minima is taken as the zero point of the mode amplitude.} \label{fig:CrossSections}
\end{figure}

Figure \ref{fig:CrossSections} shows two cross-sections of the BO surface in the plane spanned by the two soft modes of the tetrahedral structure. Figure \ref{subfig:mintomax} shows a slice running from one of the three minima of the surface to the opposite high point passing through the origin of coordinates $(u_1,u_2)=(0,0)$. Figure \ref{subfig:mintomin} shows a slice that runs from one minimum to an adjacent minimum, and this line does not cross the origin of coordinates. 

